\documentclass[apj]{emulateapj}
\usepackage{epsfig}
\newcommand{\etal}{et al.}

\newcommand\chandra{{\it Chandra}}

\newcommand\xmm{{\it XMM-Newton}}

\def\snr{Puppis~A}
\def\psr{\rm{PSR J0821$-$4300}}
\def\kespsr{\rm{PSR J1852$+$0040}}
\def\src{\rm{RX J0822$-$4300}}
\def\pks{{\rm PKS~1209$-$51/52}}
\def\pkssrc{{\rm 1E~1207.4$-$5209}}
\def\pkspsr{{PSR~J1210$-$5226}}

\def\simlt{\mathrel{\hbox{\rlap{\hbox{\lower4pt\hbox{$\sim$}}}\hbox{$<$}}}}
\def\simgt{\mathrel{\hbox{\rlap{\hbox{\lower4pt\hbox{$\sim$}}}\hbox{$>$}}}}

\slugcomment{Submitted December 29, 2008}

\shorttitle{Discovery of a 112~ms X-ray Pulsar in Puppis~A}
\shortauthors{Gotthelf \&  Halpern}

\begin{document}

\title{Discovery of a 112~ms X-ray Pulsar in Puppis~A: \\
Further Evidence of Neutron Stars Weakly Magnetized at Birth}

\author{E. V. Gotthelf \& J. P. Halpern}
\affil{Columbia Astrophysics Laboratory, Columbia University, 
New York, NY 10027-6601; \\ eric@astro.columbia.edu,
jules@astro.columbia.edu}

\begin{abstract}
We report the discovery of 112-ms X-ray pulsations from \src, the
compact central object (CCO) in the supernova remnant \snr, in two
archival {\it Newton X-Ray Multi-Mirror Mission} observations taken in
2001.  The sinusoidal light curve has a pulsed fraction of $11\%$ with
an abrupt $180^{\circ}$ change in phase at 1.2~keV.  The observed
phase shift and modulation are likely the result of emission from
opposing thermal hot spots of distinct temperatures. Phase-resolved
spectra reveal an emission feature at $E_{\rm line} = 0.8$~keV
associated with the cooler region, possibly due to an electron
cyclotron resonance effect similar to that seen in the spectrum of the
CCO pulsar \pkssrc.  No change in the spin period of
\psr\ is detected in 7 months, with a $2\sigma$ upper limit on the
period derivative of $\dot P < 8.3 \times 10^{-15}$.  This implies
limits on the spin-down energy loss rate of $\dot E < 2.3 \times
10^{35}$~erg~s$^{-1}$, the surface magnetic dipole field strength $B_s
< 9.8 \times 10^{11}$~G, and the spin-down age $\tau_c > 220$~kyr.
The latter is much longer than the SNR age, indicating that \psr\ was
born spinning near its present period. Its properties are remarkably
similar to those of the two other known CCO pulsars, demonstrating the
existence of a class of neutron stars born with weak magnetic fields
related to a slow original spin.  These results are also of importance
in understanding the extreme transverse velocity of
\psr, favoring the hydrodynamic instability mechanism in the supernova
explosion.

\end{abstract}


\section {Introduction}

Recent timing studies of central compact objects (CCOs) in supernova
remnants (SNRs) have found young neutron
stars (NSs) whose spin-down is imperceptible due to a weak dipole magnetic field
\citep{got08}. More generally, CCOs are defined by their
steady, predominantly thermal X-ray emission, lack of optical or radio
counterparts, and absence of pulsar wind nebulae (see reviews by
\citealt{pav04} and \citealt{del08}). Of the six firm members of the
CCO class, two are identified explicitly as pulsars, while searches
for pulsations from the other CCOs have been unsuccessful. In
particular, claims have been made for several candidate 
spin periods of \src, the CCO in \snr\ \citep{pav99,hui06a},
but none has been confirmed.

\nobreak \src\ has long been considered the stellar 
remnant of the supernova explosion that produced \snr,
given its central location, lack of an optical counterpart, and
hot thermal spectral properties \citep{pet82,pet96}. 
\snr\ itself is an oxygen-rich SNR, similar to Cas~A.
Its age is estimated as $3.7 \pm 0.4$~kyr from the proper motion of
its oxygen knots (Winkler \etal\ 1998), and its distance $d = 2.2 \pm
0.3$~kpc is derived from \ion{H}{1} absorption features \citep{rey95}.
More recently, an unusually large transverse velocity of $\approx
1600$~km~s$^{-1}$ was measured for \src\ \citep{win07,hui06b}, testing
proposed mechanisms of natal kick.

In this Letter we present a reanalysis of a pair of timing
observation of \src\ acquired with the {\it Newton X-Ray Multi-Mirror
Mission} (\xmm) observatory. We find a highly significant pulsed signal
in both data sets that imposes an interesting upper limit on the
period derivative.  \psr\ becomes the third CCO pulsar, with
timing and spectral properties similar to those of the
two other CCO pulsars, \kespsr\ in Kes~79 \citep{got05,hal07} and
\pkspsr\ in \pks\ \citep{got07}. We discuss the physical basis
for the class of CCO pulsars and how their natal properties differ from
radio pulsars and magnetars.

\section{XMM-Newton Observations}

Two archival \xmm\ observations of \src\
are best suited to search for coherent pulsations.
These were obtained on 2001 April 15 and
November 8 using the European Photon
Imaging Camera (EPIC; \citealt{tur03}) operating in ``small-window''
mode ($4\farcm3 \times 4\farcm3$ field-of-view). This mode provides
5.7~ms time resolution, allowing a search for even the most rapidly
rotating young pulsar. The EPIC~pn detector is sensitive to X-rays in the
nominal $0.1-12$~keV range with energy resolution $\Delta E/E
\approx 0.1/\sqrt{E({\rm keV})}$. The medium filter was used
with the target placed at the default EPIC~pn focal plane
location for a point source. The concurrent EPIC~MOS data 
are not used in this work because of its more limited (2.7~s) time
resolution.

\begin{deluxetable}{cccccc}
\tablewidth{0pt}
\tablecolumns{6}
\tablecaption{\xmm\ Timing Results for \psr}
\tablehead{
\colhead{Epoch}	& \colhead{Dur.} & \colhead{Bkgnd.\tablenotemark{a}} & \colhead{Source\tablenotemark{b}} & \colhead{Period\tablenotemark{c}} & \colhead{$f_p$\tablenotemark{d}} \\
\colhead{(MJD)} & \colhead{(s)}  & \colhead{(s$^{-1}$)}  & \colhead{(s$^{-1}$)} & \colhead{(s)} & \colhead{(\%)}
}
\startdata
 52,014.325 & 22,637 & 0.131(4) & 0.653(8) & 0.112799416(51) & 11(1) \\
 52,221.765 & 22,511 & 0.158(5) & 0.668(8) & 0.112799437(40) & 11(1) \\
\enddata
\tablecomments{Statistical uncertainty in the last digit is given in parentheses.}
\tablenotetext{a}{\footnotesize Background rate in the $1.5-4.5$ keV band
in a $0\farcm5<r<0\farcm8$ annular aperture, corrected for dead-time.}
\tablenotetext{b}{\footnotesize Count rate in the $1.5-4.5$ keV band in a
$r = 0\farcm5$ aperture, corrected for dead-time and background.}
\tablenotetext{c}{\footnotesize Period derived from a $Z^2_1$ test. Uncertainty is $1\sigma$, computed by the Monte Carlo method described in \citet{got99}.}
\tablenotetext{d}{\footnotesize Observed pulsed fraction, defined as $f_p \equiv N({\rm pulsed}) /  N({\rm total})$, after subtracting background.}
\label{timetable}
\end{deluxetable}

\begin{figure*}[t]
\centerline{
\hfill
\psfig{figure=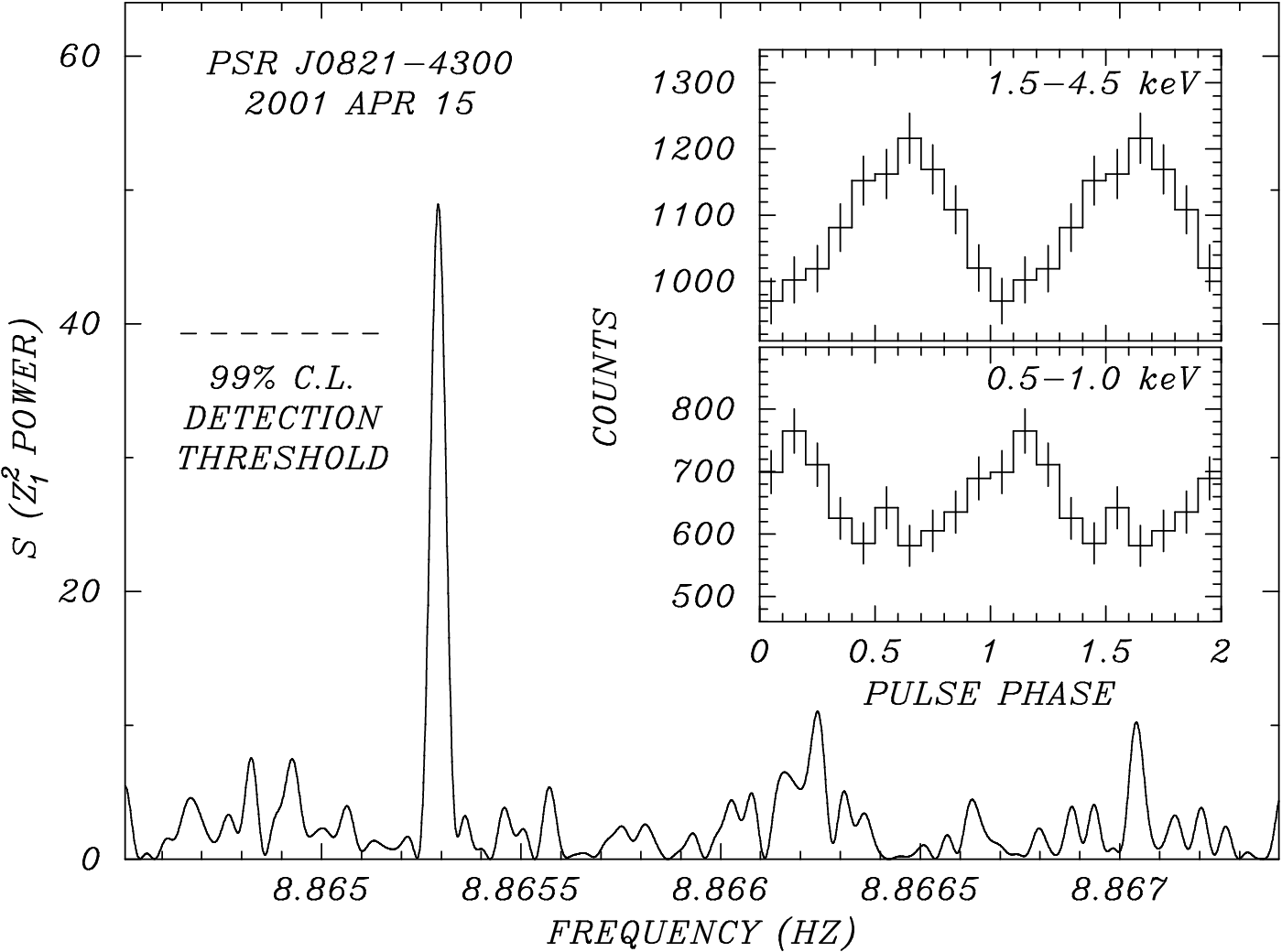,width=0.35\linewidth,angle=270}
\hfill
\psfig{figure=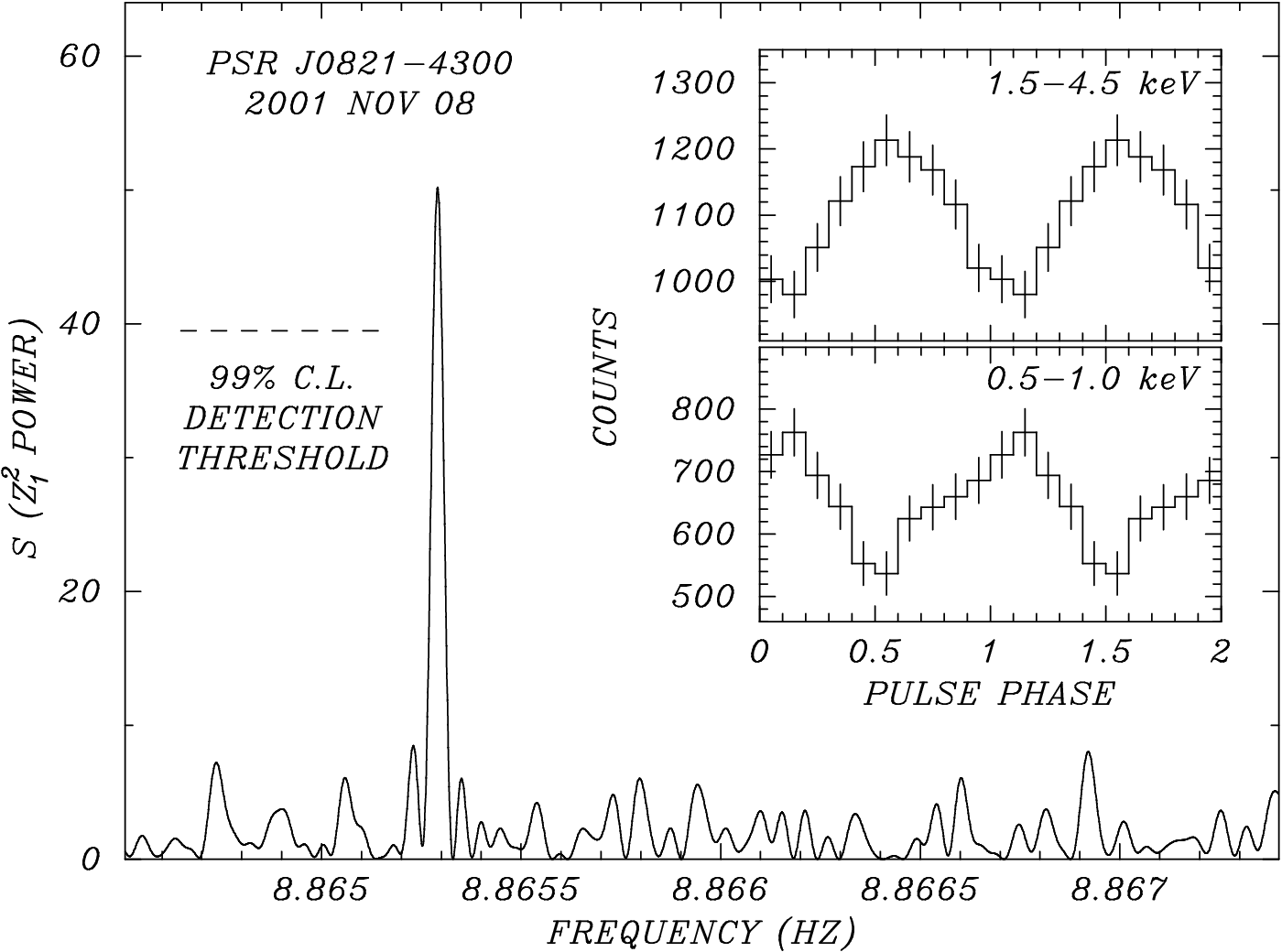,width=0.35\linewidth,angle=270}
\hfill
}
\caption{Discovery of \psr\ in \snr\ using \xmm\ EPIC~pn data acquired
on 2001 April 15 ({\it left}) and November 8 ({\it right}). The power
spectrum of $1.5-4.5$~keV photons is extracted from a
$30^{\prime\prime}$ radius aperture at the location of \src. The
detection threshold for a blind search for a period $P>12$~ms is
indicated. {\it Inset\/}: Background subtracted pulse profiles of
\psr\ in two energy bands illustrating the phase shift between the
two; the soft photons are extracted from a $18^{\prime\prime}$ radius
aperture to minimize the SNR background contamination.}
\label{timeplot}
\end{figure*}

We reprocessed the EPIC~pn data using the Science Analysis System
version SAS 7.0.0 (2006028\_1801) and screened the photon event lists
using the standard criteria. Both observations were
uncontaminated by flare events, providing $\approx
22.5$~ks of good EPIC~pn exposure time during each epoch. After taking
into account the CCD readout dead-time (29\%), this translates to
16~ks of live-time per observation in ``small window'' mode.  Photon
arrival times were converted to the solar system barycenter using
source coordinates $08^{\rm h}21^{\rm m}57\fs37,
-43\arcdeg00\arcmin17\farcs0$ (J2000.0) derived from a contemporaneous
\chandra\ observation. In the following analysis we choose a source
aperture $0\farcm5$ in radius centered on \src\ to maximize signal
over background, except where otherwise noted.  This
region encloses $\geq 80\%$ of the source flux. Count rates and
exposure times are presented in Table~\ref{timetable}.

\subsection{Timing Analysis}

To search for a pulsed signal, arrival times of photons were initially
extracted in the energy band $1.5-4.5$~keV under the hypothesis that
hard X-rays coming from a smaller area might be more strongly pulsed
than soft X-rays from the full stellar surface.  A $2^{23}$-bin Fast
Fourier Transform was used.  The most significant signal detected was
$P = 112$~ms in both EPIC~pn data sets, with no higher harmonics. We
constructed a periodogram centered on this signal using the $Z^2_1$
(Rayleigh) test \citep{buc83} and localized the pulsed emission with a
peak statistic of $Z^2_1 = 48.94$ and $Z^2_1 = 50.20$ for the 2001
April and November observations, respectively.  These statistics
correspond to 99.991\% and 99.995\% confidence, respectively, after
allowing for the number of independent trials ($2 \Delta T_{\rm span}
/ P_{\rm min}$) in a blind search for $P_{\rm min} > 12$~ms, the
Nyquist limit.  The timing results are summarized in
Table~\ref{timetable}.

The periods derived from the $Z^2_1$ test are statistically identical.
The periodogram derived from each data set is shown in
Figure~\ref{timeplot} along with the pulse profile folded at the best
period.  With a single coherent fold using both data sets, a total
power of $Z^2_1 = 95.98$ is recovered without the need for a period
derivative. This corresponds to a negligible probability of chance
occurrence. The formal result is $\dot P = (1.2
\pm 3.6) \times 10^{-15}$~s~s$^{-1}$.  The uncertainty is $1\sigma$
and is computed by propagating the uncertainties on the individual
period measurements.  From the $2\sigma$ upper-limit of $\dot P < 8.3
\times 10^{-15}$~s~s$^{-1}$ we constrain the spin-down power of
\psr\ to $\dot E \equiv 4\pi^2 I\dot P/P^3 < 2.3 \times 10^{35}$~erg~s$^{-1}$, 
the surface dipole magnetic field strength to $B_s = 3.2 \times
10^{19} \sqrt{P\dot P}< 9.8 \times 10^{11}$~G, and the characteristic
age to $\tau_c \equiv P/2\dot P > 220$~kyr.

The background-subtracted pulsed fraction is $11\pm1\%$ for both
observations.  The pulsed fraction is defined here as $f_p \equiv
N({\rm pulsed})/N({\rm total})$, where we choose the minimum of the
folded light curves as the unpulsed level.  
The pulse shape is evidently unchanged between observations,
suggesting a stable underlying emission process.  No other significant
signal is detected in either data set.  In particular, we examined the
two weaker candidate signals reported by
\citet{hui06a} from these observations.  Although we see these peaks,
neither is statistically significant or repeatable between the two
observations. We also analyzed all archival X-ray data sets that might
yield detections of \psr\ and further constrain its timing parameters,
but none proved sensitive enough.

\begin{figure}[t]
\centerline{
\hfill
\psfig{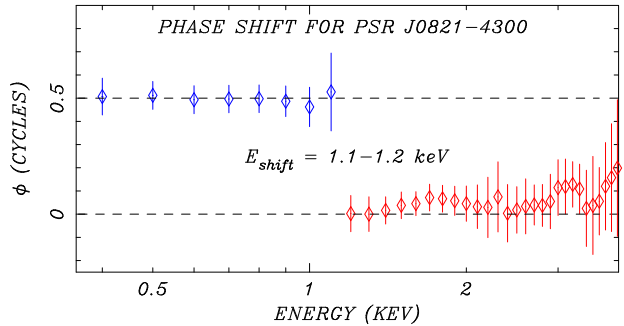}
\hfill
}
\caption{Pulse phase as a function of energy for \psr.
The phases and errors are determined by cross-correlating
energy-selected folded light curves with a master template. The phase
is computed for overlapping 0.5~keV wide energy slices incremented in
0.1~keV steps. Phase zero is defined as the minimum of the modulation
as shown in Figure~\ref{timeplot}.  The result using the April 15 data
set is shown; the November 8 data yields an equivalent result. Both
show a half cycle phase shift at $E_{\rm shift}=1.1-1.2$~keV.}
\label{phaseplot}
\end{figure}

A search for the signal in the softer energy band reveals that the
pulse is indeed present, below 1.0~keV, but shifted in phase by half a
cycle relative to the hard pulse.  Figure~\ref{phaseplot} shows that
the phase shift is remarkably abrupt, with a half cycle step ($\Delta
\phi = 0.5$) at an energy of $1.1-1.2$~keV. 
The location of the shift is unresolved to better than $\sim
0.25$~keV, limited by the counting statistics and the EPIC~pn energy
resolution of $\sim 0.1$~keV at 1~keV. This shift effectively cancels
the signal in a full-spectrum search. 

\subsection{Phase-Resolved Spectroscopy}

\citet{hui06a} performed a comprehensive spectral analysis of
the \xmm\ data on \src, showing that a two-component fit is necessary.
They preferred a double blackbody model, which is also compatible with
our expectation that surface thermal emission dominates over any other
source of X-rays from CCOs. We can investigate the cause of the phase
shift in the pulse profiles by examining pulse phase-resolved spectra
and searching for features that may be phase-dependent.  In
particular, we extracted two spectra, one in the phase band centered
on the pulse peak found in the harder energy band ($\phi=0.5$), as
shown in Figure~\ref{timeplot}, and the other centered on the phase
band found for the softer emission ($\phi=0.0$).  Each spectrum covers
a phase interval 0.4~cycles and is referred to herein as the hard and
soft phase spectra.

\begin{figure}[t]
\centerline{
\hfill
\psfig{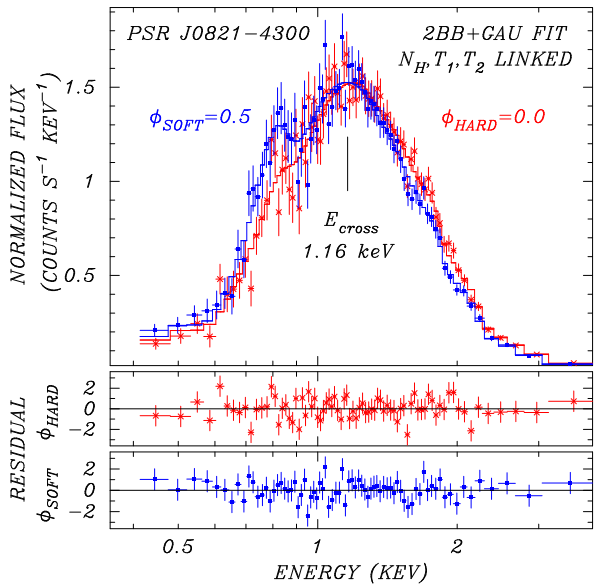}
\hfill
}
\caption{Phase-resolved spectra from the summed \xmm\ observations
of \psr. The two spectra correspond to two phase regions, 0.4 cycles
wide, one centered on the pulse peak found in the softer energy band
(``$\phi_{SOFT}$'') and the other centered on the peak found in the
harder energy band (``$\phi_{HARD}$'').  In each case, a double
blackbody model plus Gaussian emission line is fitted with the
parameters given in Table~\ref{spectab}; the column density and
temperatures are linked. The cross-over energy, $E_{\rm
cross}=1.16$~keV, of the two spectra, is shown. Residuals of the data
from the model for the two spectra are shown in the lower panels; the
soft phase data require a significant line component at $E_{\rm
line}=0.79$~keV.}
\label{specplot}
\end{figure}

We combined data from both observations and grouped the two phase
resolved spectra with a minimum of 100 counts per channel. These were
fitted to the double blackbody model in the $0.4-4.0$~keV energy
range. We find that both spectra can be fitted with the same component
temperatures, with phase-dependent normalization. The fitted spectra
are shown in Figure~\ref{specplot} and the best fit parameters are listed
in Table~\ref{spectab}. For the soft phase there is clear evidence of
positive residuals from the continuum model in the $0.7-0.9$~keV
range. This is well modeled by the addition of a Gaussian emission
line of $E_{\rm line}=0.79\pm0.02$~keV with an equivalent width of $W
\approx 55$~eV.  Comparing a fit without and with the line feature
results in a change from $\chi^2 = 69.41$ to $\chi^2 = 53.45$, for 70
and 67 degrees-of-freedom, respectively. For the hard-phase spectrum,
the continuum model provides a satisfactory fit without the need for
an added line component ($W \simlt 9$~eV).

\begin{deluxetable}{lcc}
\tablewidth{0pt}
\tablecolumns{3}
\tablecaption{Phase Resolved \xmm\ Spectral Results for \psr}
\tablehead{
\colhead{Parameter}	& \colhead{Soft Phase} & \colhead{Hard Phase}
}
\startdata
N$_{\rm H}$~(cm$^{-2}$)            & $4.8\pm0.06$ & (Linked)\\
$kT_1$~(keV)           & $0.21^{+0.05}_{-0.02}$ & (Linked)\\
$kT_2$~(keV)           & $0.42^{+0.04}_{-0.3}$ & (Linked)\\
$L_1(BB)$~(erg~s$^{-1})$ & $4.1\times10^{33}$   & $3.3\times10^{33}$\\
$L_2(BB)$~(erg~s$^{-1})$ & $2.5\times10^{33}$   & $3.2\times10^{33}$\\
$A_1(BB)$~(cm$^2$)     & $2.1\times10^{12}$   & $1.7\times10^{12}$\\
$A_2(BB)$~(cm$^2$)     & $7.8\times10^{10}$   & $9.9\times10^{10}$\\
\tableline\\
\multispan3{\hfill \hbox{Gaussian Line Component}\hfill \vspace{3pt}}\\
\tableline
$E_{\rm line}$~(keV)       & $0.79\pm0.03$    & (Linked)     \\
$\sigma_{\rm line}$~(eV)   & $53^{+33}_{-22}$ & (Linked)     \\
$W$~(eV)                   & $\approx 55$     &   $\simlt 9$ \\
\tableline
$\chi^2$~(dof)          &    \multispan2 {\hbox{125.6 (153)}}      
\enddata
\tablecomments{\footnotesize Spectral fits in the $0.4-4$~keV energy band, for two phase regions, 0.4~cycles wide, one centered on the pulse peak found in the softer energy band (``soft phase'') and the other centered on phase found in the harder energy band (``hard phase''). 
All errors are at the 90\% confidence level.}
\label{spectab}
\end{deluxetable}

Given the excellent fits to the double blackbody model the most
natural explanation of the emission from \psr\ is that of two discrete
``hot spots'' on the surface of the neutron star, corresponding to the
two blackbody temperatures and areas. As the star rotates, the hard
emission dominates over the soft component when the hotter side is
facing the viewer, while half a cycle later, the soft emission
dominates as the harder emission is now (partially) eclipsed by the
star. Based on their phase relationship, evidently $180^{\circ}$, the
two spots are on opposite sides of the NS.

Several measurements favor this explanation: 1) both the soft and hard
phase spectra can be fitted to a double blackbody model with the same
component temperatures, but phase dependent normalizations, 2) the
phase shift energy ($E_{\rm shift} = 1.1-1.2$~keV) is in agreement
with the cross-over energy between blackbody components for the above
best fit model ($E_{\rm cross} = 1.16$~keV; cf. Figure~\ref{phaseplot}
and Figure~\ref{specplot}), 3) for this model fit, the magnitude of
the pulse modulation is consistent with the change of the derived
areas for both the soft and hard phase spectra (see
Table~\ref{spectab}).

In this picture, the geometry is strongly constrained.  The small
modulation requires that the spin axis be nearly aligned with the
line-of-sight. Furthermore, the hot spot axis must be nearly
perpendicular to the spin axis since both temperature components are
evident throughout the rotation phase. Future modeling of the observed
spectrum and light curves will allow us to test this hypothesis, of
opposed thermal hot-spots for \psr, and quantify the hot-spot and
viewing geometries (e.g., \citealt{per08}). This study should also 
provide important constraints on the previously allowed spectral 
models \citep{hui06a} of the pulsar.

\section{Discussion}

\subsection{Physical Basis of the CCO Class}

In the dipole spin-down formalism, the $2\sigma$ lower limit of
$\tau_c > 220$~kyr on the characteristic age of \psr, in combination
with its true (SNR) age of $T = 3.7$~kyr, implies that its initial
spin period $P_0=P\sqrt{1-T/\tau_c}$ is identical to its current
period $P = 112$~ms to $<1\%$ accuracy.  Thus, \psr\ becomes the third
CCO whose initial spin period is known to be large, and whose surface
dipole magnetic field strength, $B_s < 9.8 \times 10^{11}$~G, is
smaller than those of the canonical young radio pulsars.  In
comparison, the upper limits on $B_s$ from timing of
\pkspsr\ \citep{got07} and \kespsr\ \citep{hal07} are
$<3.3 \times 10^{11}$~G and $<1.5 \times 10^{11}$~G, respectively.

\psr\ is virtually a twin of the
105~ms pulsar \kespsr\ in Kes~79 which, however, is $\sim 80\%$
pulsed.  In contrast, the weak ($f_p \approx 11\%$) pulse of \psr\ is
more similar to the light curve of the 424~ms pulsar \pkspsr.  The
latter is most notable for the prominent cyclotron absorption lines in
its soft X-ray spectrum. For \psr, the likely detection of a weak
emission line is reminiscent of the energy dependence of the pulse
phase in \pkspsr\ \citep{pav02,del04}, which suggests that electron
cyclotron resonance processes may be affecting its pulse shape. The
surface $B$-field strength of \pkspsr\ is estimated to be $\approx 8
\times 10^{10}$~G from its absorption features \citep{big03}, in
particular, the cyclotron fundamental at 0.7~keV, and the magnetic field of
\psr\ may be similar.

The physics of the CCO class, evidently characterized by weak magnetic
field and slow, original spin, may lie in the turbulent dynamo that
generates the magnetic field.  The strength of the dynamo depends on
the rotation rate of the proto-neutron star \citep{tho93,bon06}, so
the magnetic field strength is inversely correlated with the initial
period.

\subsection{Spin-Kick Mechanisms}

\chandra\ proper motion measurements of \src\ reveal an
extraordinarily large transverse velocity ($1570\pm 240\, d_2$
km~s$^{-1}$, \citealt{win07}; $1122 \pm 360\, d_{2.2}$ km~s$^{-1}$,
\citealt{hui06b}). As discussed by \citet{win07}, this high velocity 
restricts the possible models for NS kicks, of which there
are three broad categories (see \citealt{lai01a} and \citealt{wan06}).
The discovery of the spin parameters of \psr\ now allows us to eliminate
two of these models: 1) the magnetic-induced asymmetric neutrino emission
model requires a magnetic field of $>10^{15}$~G \citep{lai01a} and can only
produce velocities $<250$ km~s$^{-1}$, and 2) the
electromagnetic rocket effect \citep{har75,lai01a} needs an initial
spin period $<1$~ms to achieve the velocity of
\psr. Therefore, the third model, that of hydrodynamic instability
during the explosion \citep{bur07}, emerges as the most likely
explanation for large NS kicks. This is consistent with a highly
asymmetric explosion, as indicated for \snr, wherein the momentum of
the oxygen knots match the momentum of \src, but in the opposite direction
\citep{win07}.

Interestingly, the hydrodynamic mechanism is the one mechanism of the
three candidate theories that does {\it not} lead naturally to
alignment between the spin axis and the direction of proper motion
that is observed in the slowly moving Crab ($v_t \approx 120$
km~s$^{-1}$) and Vela ($v_t \approx 60$ km~s$^{-1}$) pulsars, and is
statistically evident in larger samples of pulsars \citep{jon05,ng07}.
If the spin of \psr\ were aligned with its velocity, which is nearly
in the plane of the sky, then we would have expected to see a large
rotational modulation from any surface hot spot.  On the contrary, if
the slow initial spin period of \psr\ is longer than the duration of
the kick mechanism, the impulse of the asymmetric kick could have been
expended in one direction rather than being averaged out by rotation.
Thus, the long spin period of \psr\ allows its spin axis and velocity
vector to be misaligned. This is consistent with our model for the
weak modulation, where the spin axis is close to the line-of-sight,
nearly perpendicular to the velocity vector, as is allowed for a large
kick from the hydrodynamic mechanism.

\section{Conclusions and Future Work}

The discovery of \psr\ advances the argument, reviewed by
\citet{got08}, that slow original spin and weak magnetic field constitute the
physical basis of the CCO class.  There may even be a causal
connection between slow natal spin and weak magnetic field through a
turbulent dynamo that creates the magnetic field.  The next step is a
dedicated timing study of \psr\ to determine whether it is spinning
down steadily, and to measure its period derivative and dipole
magnetic field.  An alternative possibility that it is accreting from a
fallback disk \citep{hal07} can also be tested by detailed timing.
Similar studies are underway for the other CCO pulsars.

We predict that most if not all of the remaining CCOs will turn out to
be weakly magnetized pulsars with $P \simgt 0.1$~s.  These may include
the CCO in Cassiopeia~A, and possibly an as-yet unseen NS in SN 1987A.
In the absence of spin parameters, is difficult to discriminate
between a low magnetic field CCO and a quiescent magnetar on the basis of
an X-ray spectrum alone, as these classes have similar spectral
properties.  Before the CCO pulsars were known, both Cas~A and SN
1987A were hypothesized to host magnetars.  However, the recent
retraction \citep{kim08,dwe08} of claimed evidence from the {\it
Spitzer Space Telescope} for an historic SGR-like outburst in Cas~A
\citep{kra05} leaves a weakly magnetized NS as the most compelling
model for that youngest of known CCOs.  In the case of Cas~A, we must
allow for the possibility that an energy-dependent phase shift similar
to \psr\ makes it difficult to detect pulsations. We are searching for
similar effects in all CCOs.

\acknowledgments

This investigation is based on observations obtained with \xmm, an ESA
science mission with instruments and contributions directly funded by
ESA Member States and NASA. This work is was made possible by NASA XMM 
grant NNX08AX67X, which provided critical hardware support.

\end{document}